\documentclass[runningheads]{llncs}
\usepackage{color}
\usepackage{graphicx}
\usepackage{cite}
\usepackage{hyperref}  
\usepackage{multirow}
\usepackage{amsmath,amssymb,amsfonts}

\begin{document}
\title{AirwayNet: A Voxel-Connectivity Aware Approach for Accurate Airway Segmentation Using Convolutional Neural Networks}
\titlerunning{AirwayNet}
%
\authorrunning{Y. Qin et al.}
\author{Yulei Qin\inst{1,2} \and
Mingjian Chen\inst{1,2} \and
Hao Zheng\inst{1,2} \and
Yun Gu\inst{1,2} \and
Mali Shen\inst{4} \and
Jie Yang\inst{1,2}\thanks{Corresponding author: Jie Yang. This research is partially supported by National Natural Science Foundation of China (No. 61572315, 61603248), IMR2018QY01, 973 Plan of China (No. 2015CB856004), and Committee of Science and Technology, Shanghai, China (No. 17JC1403000).} \and
Xiaolin Huang\inst{1,2} \and
Yue-Min Zhu\inst{3} \and
Guang-Zhong Yang\inst{2,4}
}

\institute{
Institute of Image Processing and Pattern Recognition,\\
Shanghai Jiao Tong University, Shanghai, China\\\email{jieyang@sjtu.edu.cn} \and
Institute of Medical Robotics, Shanghai Jiao Tong University, Shanghai, China \and
CREATIS (CNRS UMR 5220, INSERM U1206), INSA Lyon, Lyon, France \and
Hamlyn Centre for Robotic Surgery, Imperial College London, London, UK
}


%

%
\maketitle    

\begin{abstract}
Airway segmentation on CT scans is critical for pulmonary disease diagnosis and endobronchial navigation. Manual extraction of airway requires strenuous efforts due to the complicated structure and various appearance of airway. For automatic airway extraction, convolutional neural networks (CNNs) based methods have recently become the state-of-the-art approach. However, there still remains a challenge for CNNs to perceive the tree-like pattern and comprehend the connectivity of airway. To address this, we propose a voxel-connectivity aware approach named AirwayNet for accurate airway segmentation. By connectivity modeling, conventional binary segmentation task is transformed into 26 tasks of connectivity prediction. Thus, our AirwayNet learns both airway structure and relationship between neighboring voxels. To take advantage of context knowledge, lung distance map and voxel coordinates are fed into AirwayNet as additional semantic information. Compared to existing approaches, AirwayNet achieved superior performance, demonstrating the effectiveness of the network's awareness of voxel connectivity.
\end{abstract}

\section{Introduction}
Pulmonary diseases, including chronic obstructive pulmonary diseases (COPD) and lung cancer, pose high risks to human health. The standard computed tomography (CT) helps radiologists detect pathological changes. For tracheal and bronchial surgery, airway tree modeling on CT scans is often considered a prerequisite. Meticulous efforts are required to manually segment airway due to its tree-like structure and variety in size, shape, and intensity.

Several methods have been proposed for airway segmentation on CT images. Van Rikxoort et al. \cite{van2009automatic} proposed a region growing method with adaptive thresholding. Xu et al. \cite{xu2015hybrid} combined two tubular structure enhancement techniques within the fuzzy connectedness segmentation framework. Lo et al. \cite{lo2010vessel} designed a learning-based approach that models airway appearance and utilized vessel orientation similarity. In the EXACT'09 challenge, fifteen airway extraction algorithms were summarized by Lo et al. \cite{lo2012extraction}. Most algorithms adopted region growing with additional constraints such as tube likeness. Although successfully segmenting bronchi of large size, these conventional methods performed worse on peripheral bronchi.

Recently, convolutional neural networks (CNNs) were increasingly used in segmentation tasks \cite{ronneberger2015u, cciccek20163d}. For airway extraction, CNNs-based methods \cite{charbonnier2017improving, yun2019improvement, meng2017tracking, jin20173d, juarez2018automatic} were developed and proved superior to previous methods in \cite{lo2012extraction}. Charbonnier et al. \cite{charbonnier2017improving} and Yun et al. \cite{yun2019improvement} respectively used two-dimensional (2-D) and 2.5-D CNNs on already coarsely segmented bronchi to reduce false positives and increase detected tree length. Meng et al. \cite{meng2017tracking} embedded CNNs-based segmentation into the airway volume of interest (VOI) tracking framework. Jin et al. \cite{jin20173d} employed graph-based refinement on the probability output of CNNs. Juarez et al. \cite{juarez2018automatic} designed an end-to-end CNN model with simple pre- and post-processing. Despite their satisfactory performance, these methods did not specifically take airway connectivity into consideration, leaving room for improvement.

In this paper, we propose AirwayNet, a CNNs-based approach for accurate airway segmentation. Considering that the tree-like structure of airway is rather complex and the prediction of airway candidates is prone to discontinuity, we put emphasis on the connectivity of airway voxels. Unlike previous methods, we do not directly train the network to classify foreground and background voxels. Instead, binary segmentation task is transformed into 26 tasks of predicting whether a voxel is connected to its neighbors. Since airway voxels are stretching from the main bronchus towards bronchiole end as a whole connected region, we consider it a good solution to enable the model being aware of voxel connectivity. Previous work on salient segmentation \cite{kampffmeyer2019connnet} demonstrated that connectivity modeling spontaneously encodes the relation between two pixels. Therefore, we design a voxel connectivity-aware approach to better comprehend the inherent structure of airway. Our main contributions can be summarized as follows: (1) The voxel connectivity of airway is modeled using conventional binary ground-truth labels to better serve the airway segmentation task. The proposed AirwayNet automatically learns relationship between adjacent voxels and discriminates airway from the background. For each voxel, the network predicts not only its probability of being airway but also its connectivity to neighbors. (2) To effectively utilize wide-range context knowledge, voxels' coordinates and their distance to lung borders are leveraged by the model as additional semantic information. (3) Trained using 20 cases and evaluated on 10 additional cases, AirwayNet achieved the highest Dice coefficient and true positive rate of $90.2\%$, $84.7\%$, respectively, compared to state-of-the-art methods.

\section{Method}
\begin{figure}[htbp]
\centerline{\includegraphics[scale=0.4]{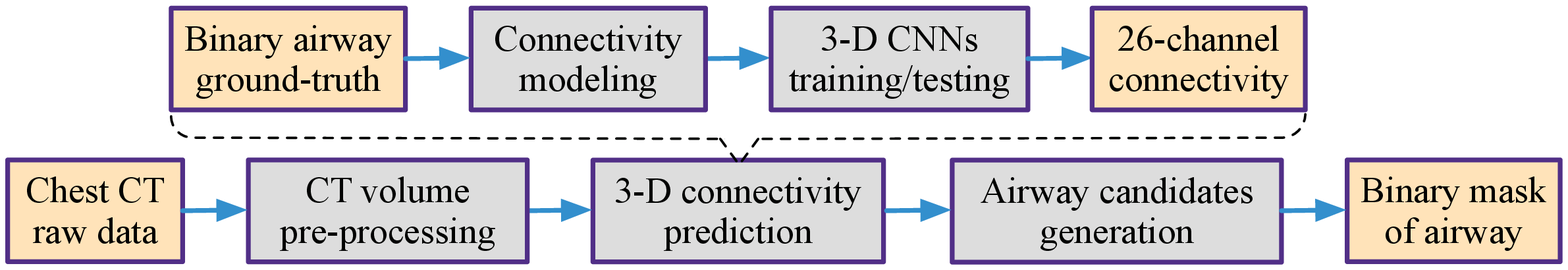}}
\caption{Flowchart of the proposed AirwayNet.}
\label{fig:flowchart}
\end{figure}

In this section, we first introduce how to model voxel connectivity and transform the segmentation problem into connectivity prediction problem. Then, details about CT pre-processing and the 3-D CNNs-based connectivity prediction are provided. Finally, we discuss the generation process of airway candidates. The flowchart of the proposed AirwayNet is depicted in Fig. \ref{fig:flowchart}.

\subsection{Connectivity Modeling Using Binary Ground-truth Labels}
\begin{figure}[htbp]
\centerline{\includegraphics[width=\columnwidth]{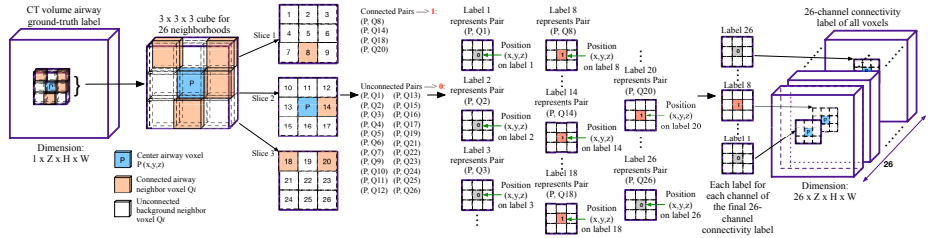}}
\caption{Illustration of 26-connectivity modeling. The binary ground-truth of airway (Dim: $1\times Z\times H\times W$) is transformed into a connectivity label (Dim: $26\times Z\times H\times W$).}
\label{fig:connectivity}
\end{figure}

In a three-dimensional (3-D) CT, 26-connectivity describes well the relation between one voxel and its 26 neighbors (see Fig. \ref{fig:connectivity}). Given a voxel $P=(x, y, z)$ and its neighbor $Q=(u, v, w)$, the distance between $P$ and $Q$ is restricted by $d(P,Q)=max(|(x-u)|, |(y-v)|, |(z-w)|)\le 1$, which means that $Q$ is located within a $3\times3\times3$ cube centered at $P$. We index neighbors $Q$ from $1$ to $26$ and denote each voxel pair $(P, Q_i), i\in \{1,2,...,26\}$ as a connectivity orientation. Each orientation is encoded using a 1-channel binary label. If both $P$ and $Q_i$ are airway voxels, then the pair $(P, Q_i)$ is connected and the corresponding position ``$P$" on the $i$-th label is marked as 1. Otherwise, we mark 0 on the $i$-th label to represent disconnected pair $(P, Q_i)$. By sliding such a $3\times3\times3$ window over each voxel, we obtain 26 binary labels and concatenate them into a 26-channel connectivity label. Zero padding is performed on CT volume borders to keep the size of generated labels unchanged. Such connectivity label encodes both ground-truth position and 26-connectivity relation between airway voxels. Note that all operations are performed on conventional binary labels of airway ground-truth. We do not require extra manual annotation for connectivity labels.

After modeling the 3-D connectivity, the original airway ground-truth label is reformed, and the conventional segmentation task is then transformed into 26 connectivity prediction tasks. The objective here is to classify and merge connected airway voxels along each connectivity orientation. An advantage of decomposing one task into 26 different tasks is that multi-task learning strategy helps the network learn more generalized features. These 26 tasks are correlated in depicting voxel connectivity, so that our AirwayNet can extract essential and robust features. Furthermore, the connectivity label emphasizes airway's structure attribute. The network trained using such label is encouraged to grasp the tree-like pattern of bronchial airway.

\subsection{CT Volume Pre-processing}
One challenge in airway segmentation is that the foreground voxels only occupy a small proportion of all CT voxels. To avoid feature learning from irrelevant parts (e.g., ribs and skin), we restrict the valid airway candidate region inside the lung area. To extract lung mask, each CT slice is first filtered with a Gaussian filter ($\sigma=1$) and binarized with a threshold ($-600$ Hounsfield unit). The connected component analysis is applied to remove unconfident candidates and the largest two components are chosen as left and right lungs, respectively. To avoid under-segmentation, we replace the lung area by its convex hull on each slice if the convex hull has 50\% more area. We also perform Euclidean distance transform on the lung mask to calculate the distance map. Each voxel on the distance map records its minimum distance to the lung border. We add such map into the network because the airway's relative position to lung border is considered anatomically meaningful. To prepare for network training, CT voxel intensity is clipped by a window $[-1000,600]$ (HU) and normalized to $[0, 255]$. Due to the limit of GPU memory, the bounding box of lungs is further cropped into smaller cubes using a sliding window technique. The cropped size is $32\times224\times224$ and the sliding stride is $[8, 56, 56]$. Such cubes include abundant context knowledge for our model to distinguish between the airway and the background.

\subsection{Connectivity Prediction Based on 3-D CNNs}
\begin{figure}[bp]
\centerline{\includegraphics[width=\columnwidth]{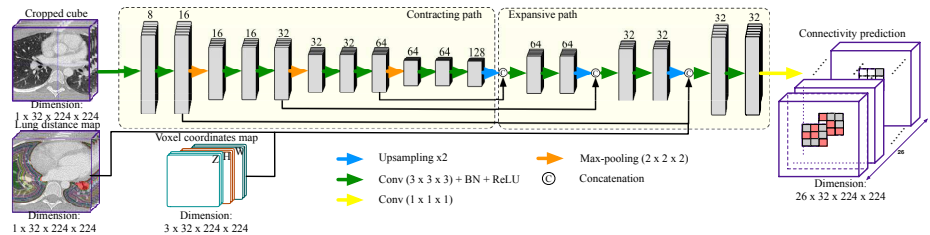}}
\caption{The architecture of the 3-D CNNs used in the proposed AirwayNet. The number of channels is denoted above each feature map.}
\label{fig:arch}
\end{figure}

Given a cropped CT cube, the objective is to use 3-D CNNs to predict voxel connectivity of the cube. The proposed AirwayNet (see Fig. \ref{fig:arch}) is based on the U-Net \cite{cciccek20163d} backbone. Our full 3-D architecture captures more spatial information than the 2-D or 2.5-D CNNs used in \cite{charbonnier2017improving,yun2019improvement} and is more suitable for learning the bronchial continuity and branching patterns. The AirwayNet consists of a contracting path and an expansive path with four resolution scales. At each resolution scale, the contracting path has two convolution layers (Conv) with batch normalization (BN) and rectified linear unit (ReLU), followed by a max-pooling layer. In the expansive path, finer features from lower resolution scale are linearly upsampled first and then concatenated with coarse features from skip connection to preserve details of thin bronchi. Since airway voxels are distributed within the large thoracic cavity, extra semantic information other than grayscale intensity is considered beneficial for the model to classify airway voxels. Here we use voxel coordinates and lung distance map, and concatenate them with features on the expansive path at the last scale. The sigmoid function is applied on the predicted connectivity cube to obtain probability distribution. We use the Dice similarity coefficient (DSC) loss to optimize our AirwayNet. For each voxel $x$ in the cropped cube $X$, given its label $y_{i}(x)$ and prediction probability $p_{i}(x), i\in\{1,2,...,26\}$, the total connectivity loss is defined as the averaged DSC loss over all channels:
\begin{equation}
\mathcal{L} = 1 - \frac{1}{26}\sum^{26}_{i=1}\frac{2\sum_{x\in X}p_i(x)y_i(x)}{\sum_{x\in X}(p_i(x)+y_i(x))+\epsilon},
\label{eq:diceloss}
\end{equation}
\noindent where $\epsilon$ is used to avoid division by zero.
\subsection{Airway Candidates Generation}
The final step is to generate airway candidates based on the predicted connectivity cube. First, a threshold $t=0.5$ is used to binarize prediction results. Here we consider that pairwise voxels should agree with each other in connectivity. For example, if voxel $P$ is connected to its neighbor $Q_{14}$ (see Fig. \ref{fig:connectivity}), then voxel P on the $14$-th connectivity channel is marked as $1$. Meanwhile, on the $13$-th channel, voxel $Q_{14}$ is marked as $1$ because $P$ is also at the position ``$Q_{13}$" of the $3\times3\times3$ neighborhood of $Q_{14}$. The connectivity between $P$ and $Q_{14}$ is coded in both the $13$-th and the $14$-th channels. Therefore, we only keep voxels that comply with such pairwise agreement. Then, channel-wise summation is performed on the connectivity cube and those non-zero voxels are marked as airway candidates. The segmentation results are multiplied with the lung field mask to filter out false positives. Finally, we follow \cite{jin20173d} to apply fuzzy connectedness segmentation on the generated candidates to consolidate the bronchial distal ends.

\section{Experiments and Results}
The experiment dataset contains 30 chest CT scans. All CT axial slices have the same size of 512$\times$512, with a pixel spacing ranging from 0.5 to 0.781 mm. The slice thickness varies from 0.5 to 1.0 mm. These scans were acquired using different scanners, protocols, dose level, and reconstruction kernels. The dataset contains cases from both healthy volunteers and patients with severe pulmonary diseases (e.g., emphysema and pneumonia). The ground-truth labels of airway were annotated and screened by well-trained experts for verification. We randomly chose 20 CT scans for training and fine-tuning hyper-parameters. Our approach was evaluated on the remaining 10 scans.

During training, cropped cubes were augmented on-the-fly via random horizontal flipping, with a probability of 0.5. We densely sampled cubes near airway region and discarded cubes near lung border randomly. This results in around 5000 training samples for each epoch. The Adam optimizer ($\beta_1=0.5, \beta_2=0.999$) was used and the learning rate was set to $10^{-4}$. The training converged after 15 epochs. Our AirwayNet was implemented in Keras with 4 NVIDIA Titan Xp GPUs. During testing, the stride of the sliding window was $[16, 128, 128]$ and the prediction results were averaged on overlapping margins. Four evaluation metrics were used: (a) Dice coefficient (DSC), (b) True positive rate (TPR), (c) False positive rate (FPR), and (d) Positive predictive value (PPV).

\begin{figure}[t]
\centerline{\includegraphics[width=\columnwidth]{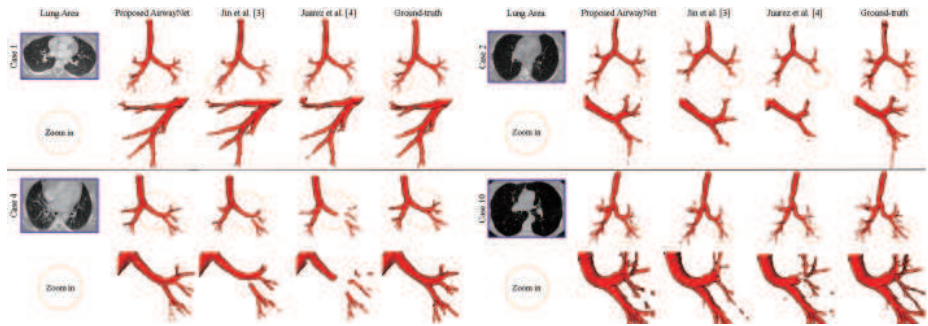}}
\caption{Comparison of airway segmentation results. Local bronchial branches are highlighted with \textbf{circles} and zoomed in to better see performance difference.}
\label{fig:comp}
\end{figure}

We compared our AirwayNet with two state-of-the-art approaches, Jin's method \cite{jin20173d} and Juarez's method \cite{juarez2018automatic}. They both employ 3-D CNNs with a sliding window technique for airway extraction. In Table \ref{table:comp1}, it is observed that our method achieved the highest DSC and TPR of 90.2$\%$, 84.7$\%$, respectively, with the comparable FPR of 0.011$\%$ and PPV of 96.6$\%$. The smallest standard deviation of DSC and TPR was achieved by AirwayNet, confirming its robustness. With lower DSC, TPR, and FPR but higher PPV, Juarez et al. \cite{juarez2018automatic} segmented more conservatively than AirwayNet and the proposed method slightly tended to classify background voxels as airway. But our TPR is 6.4\% higher than theirs on average, meaning that more airway voxels were detected by AirwayNet. In addition, qualitative results of the proposed AirwayNet are shown in Fig. \ref{fig:comp}, together with its comparison with \cite{jin20173d, juarez2018automatic}. Owing to connectivity modeling, our method enriched details of segmented peripheral bronchi. In contrast, more holes and cracks appear on the predicted bronchial airway of other methods.
\begin{table}[tbp]
\centering
\caption{Comparison of airway segmentation results (\%) on 10 chest CT scans.}
\label{table:comp1}
\scalebox{0.62}{
\begin{tabular}{|l|l|l|l|l|l|l|l|l|l|l|l|l|}
\hline
\multirow{2}{*}{Case} & \multicolumn{4}{l|}{AirwayNet} & \multicolumn{4}{l|}{Jin et al. \cite{jin20173d}} & \multicolumn{4}{l|}{Juarez et al. \cite{juarez2018automatic}} \\ \cline{2-13} 
 & DSC & TPR & FPR & PPV & DSC & TPR & FPR & PPV & DSC  & TPR & FPR & PPV \\ \hline
1 &92.4  &86.5  &0.003  &99.2  &90.4  &84.4  &0.010 &97.2  &91.0  &83.8  &0.002  &99.5  \\ \hline
2 &87.5  &78.5 &0.003  &98.8  &73.8  &60.8  &0.011  &94.0  &76.9  &62.6  &0.000  &99.8  \\ \hline
3 &91.1  &85.3  &0.004  &97.7  &88.9  &83.1  &0.008  &95.6  &91.1  &84.6  &0.002  &98.8  \\ \hline
4 &82.9  &73.1  &0.009  &95.7  &79.6  &66.5  &0.002  &99.1  &70.8  &55.6  &0.004  &97.4  \\ \hline
5 &90.8  &86.2  &0.015  &95.9  &90.6  &85.3  &0.012  &96.6  &89.3  &81.4  &0.003  &99.1  \\ \hline
6 &91.5  &89.2  &0.025  &94.0  &84.5  &88.3  &0.091  &81.0  &92.3  &87.9  &0.012  &97.1  \\ \hline
7 &90.6  &87.6  &0.025  &93.7  &88.2  &85.3  &0.034  &91.2  &90.3  &85.1  &0.014  &96.2  \\ \hline
8 &91.7  &88.9  &0.017  &94.7  &86.4  &88.1  &0.054  &84.8  &90.2  &86.5  &0.018  &94.2  \\ \hline
9 &93.1  &88.8  &0.009  &97.8  &90.5  &84.3  &0.009  &97.7  &85.9  &75.9  &0.004  &98.9  \\ \hline
10 &90.4  &83.2  &0.003  &98.9  &88.2  &79.3  &0.002  &99.4  &88.3  &79.4  &0.001  &99.5  \\ \hline
Mean &\textbf{90.2}  &\textbf{84.7}  &0.011  &96.6  &86.1  &80.5  &0.023  &93.7  &86.6  &78.3  &\textbf{0.006}  &\textbf{98.1}  \\ \hline
Std. &\textbf{2.8} &\textbf{4.9} &0.008 &1.9 &5.2 &8.9 &0.027 &5.9 &6.7 &10.3 &\textbf{0.006} &\textbf{1.7} \\ \hline
\end{tabular}}
\end{table}

We also conducted ablation study (see Table \ref{table:abstudy}) to validate whether each component of the proposed method is useful. Compared to AirwayNet without Conn and D\&C, the slightly higher FPR and lower PPV of AirwayNet may be due to the introduction of connectivity modeling and semantic information, making the model more sensitive and perceptive to airway than background. Meanwhile, the improvement in DSC and TPR proves that it is worthwhile to enable the model being aware of connectivity and semantic knowledge. Furthermore, the application of fuzzy connectedness segmentation links the tiny bronchial ends together, improving the DSC and TPR by 0.1\%, 0.3\% on average, respectively. These three components do contribute to the performance improvement.
\begin{table}[bp]
\centering
\caption{Results (\%) of ablation study of AirwayNet (mean$\pm$standard deviation). The Conn, D\&C, and FCS represent connectivity modeling, distance map \& coordinates, and fuzzy connectedness segmentation, respectively. Note that without Conn, the task of AirwayNet becomes a common segmentation task and the network is directly trained to output binary masks of airway.}
\label{table:abstudy}
\scalebox{0.62}{
\begin{tabular}{|l|l|l|l|l|}
\hline
Methods & DSC & TPR & FPR & PPV \\ \hline
AirwayNet w/o Conn &87.1$\pm$6.4  &79.3$\pm$9.6  &\textbf{0.007$\pm$0.008}  &\textbf{97.6$\pm$2.5}  \\ \hline
AirwayNet w/o D\&C &88.4$\pm$5.4  &81.6$\pm$8.7  &0.009$\pm$0.009  &97.1$\pm$2.1  \\ \hline
AirwayNet w/o FCS &90.1$\pm$2.8 &84.4$\pm$4.9 &0.011$\pm$0.008 &96.8$\pm$1.9 \\ \hline
AirwayNet &\textbf{90.2$\pm$2.8}  &\textbf{84.7$\pm$4.9}  &0.011$\pm$0.008  &96.6$\pm$1.9  \\ \hline
\end{tabular}}
\end{table}

According to Table \ref{table:comp1} and Fig. \ref{fig:comp}, our AirwayNet clearly improves the connectivity and shape reconstruction of airway but is not yet 100\% perfect, mainly due to the fact that the used training samples did not reflect all the situations including weak intensity contrast between airway lumen and wall, blurring from low quality imaging, and pathological changes of pulmonary diseases.

\section{Conclusion}
This paper proposed a CNNs-based approach, AirwayNet, for airway segmentation on CT scans. It explicitly learns voxel connectivity to perceive airway's inherent structure. By connectivity modeling, conventional segmentation task is transformed into 26 tasks of connectivity prediction, with each task classifying airway voxels along a certain connectivity orientation. The lung distance map and voxel coordinates are introduced as additional semantic information to better utilize context knowledge. Experimental results corroborated the two main strengths of AirwayNet: (1) the model's attention paid on the connectivity of airway structure and (2) the extra knowledge about voxels' position and their distance to lung borders. Consequently, the performance improvement is achieved on peripheral airway segmentation. In the future, the proposed method could further be improved by working on (1) the adoption of generative adversarial networks to produce various training samples to improve robustness on unhealthy patients' scans and (2) the exploration of specific enhancement mechanisms for thin bronchus details in low quality CT scans to improve performance.

%
%


%
%
%
\bibliographystyle{splncs04}
\bibliography{paper1008}

\begin{thebibliography}{10}
\providecommand{\url}[1]{\texttt{#1}}
\providecommand{\urlprefix}{URL }
\providecommand{\doi}[1]{https://doi.org/#1}

\bibitem{charbonnier2017improving}
Charbonnier, J.P., Van~Rikxoort, E.M., Setio, A.A., Schaefer-Prokop, C.M., van
  Ginneken, B., Ciompi, F.: Improving airway segmentation in computed
  tomography using leak detection with convolutional networks. MedIA
  \textbf{36},  52--60 (2017)

\bibitem{cciccek20163d}
{\c{C}}i{\c{c}}ek, {\"O}., Abdulkadir, A., Lienkamp, S.S., Brox, T.,
  Ronneberger, O.: {3D} {U-N}et: learning dense volumetric segmentation from
  sparse annotation. In: MICCAI. pp. 424--432. Springer (2016)

\bibitem{jin20173d}
Jin, D., Xu, Z., Harrison, A.P., George, K., Mollura, D.J.: {3D} convolutional
  neural networks with graph refinement for airway segmentation using
  incomplete data labels. In: MLMI. pp. 141--149. Springer (2017)

\bibitem{juarez2018automatic}
Juarez, A.G.U., Tiddens, H., de~Bruijne, M.: Automatic airway segmentation in
  chest {CT} using convolutional neural networks. In: Image Analysis for Moving
  Organ, Breast, and Thoracic Images, pp. 238--250. Springer (2018)

\bibitem{kampffmeyer2019connnet}
Kampffmeyer, M., Dong, N., Liang, X., Zhang, Y., Xing, E.P.: Connnet: A
  long-range relation-aware pixel-connectivity network for salient
  segmentation. IEEE TIP  \textbf{28}(5),  2518--2529 (2019)

\bibitem{lo2010vessel}
Lo, P., Sporring, J., Ashraf, H., Pedersen, J.J., de~Bruijne, M.: Vessel-guided
  airway tree segmentation: A voxel classification approach. MedIA
  \textbf{14}(4),  527--538 (2010)

\bibitem{lo2012extraction}
Lo, P., Van~Ginneken, B., Reinhardt, J.M., Yavarna, T., De~Jong, P.A., Irving,
  B., Fetita, C., Ortner, M., Pinho, R., Sijbers, J., et~al.: Extraction of
  airways from {CT} (exact'09). IEEE TMI  \textbf{31}(11),  2093--2107 (2012)

\bibitem{meng2017tracking}
Meng, Q., Roth, H.R., Kitasaka, T., Oda, M., Ueno, J., Mori, K.: Tracking and
  segmentation of the airways in chest {CT} using a fully convolutional
  network. In: MICCAI. pp. 198--207. Springer (2017)

\bibitem{ronneberger2015u}
Ronneberger, O., Fischer, P., Brox, T.: {U-N}et: Convolutional networks for
  biomedical image segmentation. In: MICCAI. pp. 234--241. Springer (2015)

\bibitem{van2009automatic}
Van~Rikxoort, E.M., Baggerman, W., van Ginneken, B.: Automatic segmentation of
  the airway tree from thoracic {CT} scans using a multi-threshold approach.
  In: Proc. of Second International Workshop on Pulmonary Image Analysis. pp.
  341--349 (2009)

\bibitem{xu2015hybrid}
Xu, Z., Bagci, U., Foster, B., Mansoor, A., Udupa, J.K., Mollura, D.J.: A
  hybrid method for airway segmentation and automated measurement of bronchial
  wall thickness on {CT}. MedIA  \textbf{24}(1),  1--17 (2015)

\bibitem{yun2019improvement}
Yun, J., Park, J., Yu, D., Yi, J., Lee, M., Park, H.J., Lee, J.G., Seo, J.B.,
  Kim, N.: Improvement of fully automated airway segmentation on volumetric
  computed tomographic images using a 2.5 dimensional convolutional neural net.
  MedIA  \textbf{51},  13--20 (2019)

\end{thebibliography}

\end{document}